\documentclass[conference]{IEEEtran}
\IEEEoverridecommandlockouts
\usepackage{cite}
\usepackage{amsmath,amssymb,amsfonts}
\usepackage{algorithmic}
\usepackage{graphicx}
\usepackage{textcomp}
\usepackage{xcolor}
\usepackage{multirow}
\usepackage{bm}  
\usepackage{url}

\DeclareMathOperator*{\argmin}{arg\,min}
\newcommand{\etal}{\textit{et al}.~}
\newcommand{\ie}{\textit{i}.\textit{e}.~}

\DeclareRobustCommand*{\IEEEauthorrefmark}[1]{%
  \raisebox{0pt}[0pt][0pt]{\textsuperscript{\footnotesize #1}}%
}

\def\BibTeX{{\rm B\kern-.05em{\sc i\kern-.025em b}\kern-.08em
    T\kern-.1667em\lower.7ex\hbox{E}\kern-.125emX}}
\begin{document}

\title{Cold Item Integration in Deep Hybrid Recommenders via Tunable Stochastic Gates}

\author{
\vspace{2mm}
    \IEEEauthorblockN{Oren~Barkan\IEEEauthorrefmark{1}\textsuperscript{,}\IEEEauthorrefmark{5}\textsuperscript{,}\textsuperscript{*}\thanks{*Equal contribution},
    Roy~Hirsch\IEEEauthorrefmark{2}\textsuperscript{,}\IEEEauthorrefmark{5}\textsuperscript{,}\textsuperscript{*},
    Ori~Katz\IEEEauthorrefmark{3}\textsuperscript{,}\IEEEauthorrefmark{5}\textsuperscript{,}\textsuperscript{*},
    Avi~Caciularu\IEEEauthorrefmark{4}, \\ Jonathan~Weill\IEEEauthorrefmark{5},  Noam~Koenigstein\IEEEauthorrefmark{2}\textsuperscript{,}\IEEEauthorrefmark{5}}
    \\
    \vspace{-1mm}
    \IEEEauthorblockA{\IEEEauthorrefmark{1}The Open University \\ 
    \IEEEauthorrefmark{2}Tel-Aviv University \\
    \IEEEauthorrefmark{3}Technion \\
    \IEEEauthorrefmark{4}Bar-Ilan University \\ \IEEEauthorrefmark{5}Microsoft}
 }

\maketitle

\begin{abstract}
  A major challenge in collaborative filtering methods is how to produce recommendations for cold items (items with no ratings), or integrate cold item into an existing catalog. Over the years, a variety of hybrid recommendation models have been proposed to address this problem by utilizing items' metadata and content along with their ratings or usage patterns.
In this work, we wish to revisit the cold start problem in order to draw attention to an overlooked challenge: the ability to integrate and balance between (regular) warm items and completely cold items.
In this case, two different challenges arise: (1) preserving high quality performance on warm items, while (2) learning to promote cold items to relevant users.
First, we show that these two objectives are in fact conflicting, and the balance between them depends on the business needs and the application at hand.
Next, we propose a novel hybrid recommendation algorithm that bridges these two conflicting objectives and enables a harmonized balance between preserving high accuracy for warm items while effectively promoting completely cold items.
We demonstrate the effectiveness of the proposed algorithm on movies, apps, and articles recommendations, and provide an empirical analysis of the cold-warm trade-off.
\end{abstract}


\section{Introduction}
\label{sec:introNrelated}
Recommender systems models are generally categorized into three main categories: (1) \emph{Collaborative Filtering (CF)} models \cite{salakhutdinov2008bayesian,koren2009matrix,barkan2016item2vec,barkan2020attentive,lavee2019actions,anchors,barkan2020explainable} that are based on ratings or implicit usage information, (2) \emph{Content Based (CB)} models that utilize items' content and metadata~\cite{malkiel2020recobert}, and (3) \emph{hybrid} models that combine both information sources \cite{wang2011collaborative,nam}. 
Historically, CF models have performed better than CB models in many well-known competitions, e.g., the well known Netflix Prize \cite{NetflixPrize} and the Yahoo! Music Challenge \cite{KDDCup11}. 
Hence, in the presence of ratings or implicit usage data, CF models are usually considered more accurate than CB models \cite{RecSys_Book}. 
However, CF models are known to suffer from the \emph{cold start} problem when it comes to recommending items with little or no ratings or implicit usage data.

In the past, utilizing certain types of items' content such as textual descriptions and images required employing challenging processing techniques. Hence, hybrid models were very limited in the types of content data they could employ.
However, recent advancements in deep learning methods gave rise to a new family of deep hybrid recommender systems that can utilize almost any type of digitally available content data  \cite{wei2017collaborative}.
By incorporating items' content together with ratings or implicit usage data, hybrid models employ a dual objective: mitigating the cold-start problem while simultaneously utilizing the items' content to improve warm item representations. 
As such, modern hybrid models appear to be the ultimate approach. 
Yet, this dual objective pose a challenge: the model strives to adhere to the CF objective and preserve accurate recommendations for warm items, while simultaneously promoting cold items to the right users.

When warm items are considered, hybrid models utilize the content data as additional ``side information'' that aids and improves the CF representation. 
However, when completely cold items are considered (items with no usage at all), the content data needs to ``take over'' and replace the missing usage representation.

Alas, existing hybrid models are not optimized to deal with completely cold items effectively. In fact, as we show later, without proper treatment of completely cold items during the training phase, a hybrid model would still attempt to utilize the content information as a mere ``correction'' over the missing CF representation that in the case of completely cold items, simply does not exist. 
In other words, in the absence of the usage data, the content representation is not optimized to completely replace (overtake) the missing CF representation and the final result is sub-optimal.

In this paper, we consider a common scenario in which a recommender system is expected to perform well both on warm items as well as on completely cold items that should be promoted to relevant users.
The trade-off between these two objectives has a major significance in real-world commercial systems where new items are regularly introduced to an existing system in order to be exposed to the right audience. 
In fact, the promotion and introduction of new (cold) items is a critical stage that often has significant and lingering effects on the product life cycle and overall consequential sales \cite{PLC1,PLC2}. 
However, as we show next, when optimizing a machine learning based recommender system, these two objectives are contradicting and this trade-off was mostly overlooked in the literature.

Ideally, a hybrid recommender would perform well on all items, and the proper balance between cold and warm item modeling would be determined based on the application at hand and the business needs e.g.,~ the proportion between the warm and cold items in the catalog, or the desired amount of exposure or promotion for cold items.
To this end, we present the Cold-Warm Harmonization (CWH) model - a novel deep hybrid model with an emphasis on harmonizing cold and warm item recommendations.
Unlike previous hybrid models, that utilize items content in order to improve warm item recommendations with very limited contribution to cold items, CWH directly addresses the items' ``cold start'' problem while supporting a tunable balance between warm and cold item modeling. 

In common with previous works, CWH learns hybrid item representations that utilize both usage (CF) data and content data in a joint manner. However, for cold items, CWH utilizes a novel dual content based representation: The first representation is similar to the one used for warm items, while the second CB representation is dedicated to cope with cold items only, and compensates for the missing CF representation. 
Additionally, a stochastic tunable gate (modeled as an observed Bernoulli variable) determines the state of each item as either \emph{warm} or \emph{cold} to enable artificially injecting cold items, hence simulating cold start scenarios, during training. 

This unique architecture alleviates the aforementioned conflicting roles of the CB representation and allows for optimal utilization of items' content.    
By introducing both cold and warm items during the training phase, the model is forced to adapt to both cases - effectively coping with completely cold items while still maintaining good performance on the warm items. In other words, CWH simulates cold-start scenarios during training time, hence compelling the model to cope with both objectives simultaneously, and in a controlled manner.

The remainder of this manuscript is organized as follows: In Section~\ref{related-Work}, we cover related work and focus specifically on earlier attempts to balance cold and warm item modeling. In Section~\ref{sec:model}, we describe the CWH model in detail. Then, in Section~\ref{sec:results} we provide evaluations and experimental results that demonstrate the effectiveness of CWH both as a hybrid recommender system as well as an effective solution for cold items when no ratings or implicit usage is available. Finally, we provide final words and conclusions in Section~\ref{sec:conclusions}.

\section{Related Work}
\label{related-Work}
The cold start problem is an active research field in the recommender systems community \cite{RecSys_Book,schein2002methods,lika2014facing,lam2008addressing,10.1145/3240323.3240404,barkan2020cold,cb2cf} and hybrid models have been studied extensively in the literature \cite{burke2002hybrid,ccano2017hybrid,wei2017collaborative,barkan2021cold,wang2011collaborative,braunhofer2014hybridisation,nam,ben2017groove,ginzburg2021self,malkiel2020optimizing}. However, the problem of balancing cold and warm item modeling in hybrid recommenders (as explained in Section~\ref{sec:introNrelated}) has been mostly overlooked.
While we cannot cover the plethora of related work on the cold-start problem or hybrid models in general, in what follows, we relate our work to the latest state-of-the-art hybrid models as well as a few earlier works that touch upon the problem of cold-vs-warm trade-off.

Collaborative Topic Regression (CTR) \cite{wang2011collaborative} and Collaborative Deep Learning (CDL) \cite{wang2015collaborative} are two highly popular hybrid models that combine a CF model based on Matrix Factorization (MF)  \cite{koren2009matrix} together with Latent Dirichlet Allocation (LDA) \cite{blei2003latent} or stacked denoising autoencoders \cite{vincent2010stacked}, respectively. The purpose of the CB objective is to support cold items when usage data is unavailable. Both CTR and CDL tackle the warm-vs-cold trade-off by employing a hyperparameter to control the importance of the MF objective with respect to the CB objective. 
In \cite{wang2015collaborative}, Wang et al. compared both CTR and CDL. As seen in Figure~4 in \cite{wang2011collaborative}, tuning this parameter has a negligible impact on cold items for both models. 
Moreover, CTR's use of the LDA model for utilizing textual information is rather anachronistic compared to modern day approaches that are usually based on word embeddings and transformers \cite{wolf2019huggingface,devlin,RecoBERT}.
Hence, we conclude that while both models do tackle the balancing problem, these models are mostly focused on warm items, and do not fully exploit the CB information for completely cold items.

Collaborative Variational Autoencoder (CVAE) \cite{cvae} is a hybrid model that was recently shown to outperform CDL, CTR as well as several other state-of-the-art baselines. CVAE employs a generative model for the CB data which is combined with learning a CF task based on implicit relationships between users and items.
In contrast to the model in this paper, CVAE does not provide any mechanism to simulate a completely cold start scenario during training.

Other deep hybrid recommender systems include the multimedia model for movie recommendations~\cite{deldjoo2018using}, HybridSVD~\cite{hybridsvd}, explainable hybrid systems~\cite{dualdiv} and content-collaborative disentanglement representation learning \cite{zhang2020content}. However, these works do not propose any way to deal with the cold-warm trade-off in hybrid models.

Recently, CB2CF \cite{cb2cf} - a deep content-to-collaborative filtering model was introduced for completely cold item recommendations. 
CB2CF works in two stages: First, CF representations are learned based on usage data. Then, the CB2CF model is applied to learn a mapping from warm items' content into their latent CF representations. This mapping is later employed in order to embed cold items into an existing CF model. 
In contrast to hybrid models, CB2CF is focused solely on completely cold items (items with zero ratings or implicit usage data) while disregarding warm items. As such, CB2CF was shown to outperform existing hybrid models on the problem of modeling completely cold items. 
However, during inference time, the warm items rely on their pre-existing CF representations which are unaware of the additional representations injected by CB2CF to completely cold items. This type of discrepancy poses a challenge to real-world systems that need to address both warm and cold items simultaneously. 

In this work, we present CWH, an end-to-end hybrid model that addresses a problem not discussed by the aforementioned works -  balancing two conflicting objectives: learning warm and cold item representations in a single unified recommender system.
By employing an effective stochastic (yet controlled) gate, we inject \emph{fake} cold items during the training phase, to force the model to adapt to both cold and warm items, simultaneously. 
Moreover, CWH employs a dual CB representation for cold items that compensate for the absence of the CF representation and alleviates the aforementioned conflicting roles of the content data. 
Consequently, CWH improves upon the state-of-the-art in several facets: (1) A novel hybrid recommender that is capable of effectively handling both warm and cold items, simultaneously. (2) A unified training procedure that improves accuracy and (3) A novel framework to balance between warm and cold item learning.
\vspace{4mm}

\section{The Cold-Warm Harmonization (CWH) Model}
\label{sec:model}
In this section, we formulate the problem setup and the CWH model. Let $\mathcal{I}=\{i\}_{i=1}^{N_u}$ and $\mathcal{J}=\{j\}_{j=1}^{N_v}$ be sets that index $N_u$ users and $N_v$ items, respectively. In addition, we assume that each item $j$ is associated with $N_c$ types of content (information sources), $\mathbf{X}_j=\{x^k_j\}_{k=1}^{N_c}$, where $x^k_j\in\mathcal{C}^k$ represent item $j$'s $k$th information source. For example, $\mathcal{C}^2$ can be images (a visual signal), and $\mathcal{C}^5$ can be the textual descriptions (unstructured text). 
The overall available content for the entire set of items is denoted by $\mathbf{X}=\{\mathbf{X}_j \}_{j=1}^{N_v}$. 
We denote the set of user-item interactions (i.e., the CF relations) by $I_y=\{(i,j)| \text{user } i \text{ consumed item } j\}$. In addition, we define $\mathbf{Y}=\{y_{ij}|(i,j)\in\mathcal{I}\times\mathcal{J}\}$, where $y_{ij}$ is a two-point observed random variable s.t. $y_{ij}=1$ if $(i,j)\in I_y$, and $y_{ij}=-1$ otherwise. Namely, $y_{ij}$ indicates whether the user $i$ consumed the item $j$ or not.
\vspace{2mm}
\subsection{A Deep Hybrid Recommender System}
\label{subsec:themodel}
Let $f^{\theta_k}:\mathcal{C}^k\rightarrow\mathbb{R}^{d_k}$ be a \emph{content analyzer} function (parameterized by $\theta_k$) that maps $x\in\mathcal{C}^k$ to a $d_k$-dimensional vector $f^{\theta_k}(x)$. For example, $f^{\theta_k}$ may be a deep neural network that analyzes the item's textual description (e.g., BERT \cite{devlin2018bert,malkiel2020optimizing}) or visual content (e.g., ResNet \cite{he2016deep}), and encodes it as a $d_k$-dimensional vector.
The unobserved parameters $\theta_k$ are learned during the model's training phase (in practice, we use a pretrained network as a backbone model, and may extend it with subsequent layers as necessary). For simplicity, we denote $f^k_j\triangleq f^{\theta_k}(x^k_j)$, which stands for the application of the content analyzer $f^{\theta_k}$ to the content information of type $k$ that is associated with the item $j$. In addition, we collectively denote $\theta_{CB}=\{\theta_1,...,\theta_{N_c}\}$.

Let $\phi^{\theta_{\phi}}:\mathbb{R}^{d_\phi}\rightarrow\mathbb{R}^{d}$, where $d_{\phi}=\sum_{k=1}^{N_c}d_k$, be a \emph{multiview} content analyzer that receives the concatenated multiview representation  $f_j = [f^1_j,...,f^{N_c}_j]$ and outputs the following $d$-dimensional vector: \begin{equation}
    \phi_j\triangleq\phi^{\theta_{\phi}}(f_j).
\end{equation}
Therefore, $\phi_j$ encodes all types of content that are associated with item $j$. In our implementation, we set $\phi^{\theta_{\phi}}$ to be a fully connected neural network with a single ReLU activated hidden layer.

Let $\mathbf{U}=\{u_i\}_{i=1}^{N_u}$ and $\mathbf{V}=\{v_j\}_{j=1}^{N_v}$ be the unobserved user and item CF representations ($\mathbf{U,V}\subset\mathbb{R}^d$). In order to score the affinity between user $i$ and item $j$, we define a neural scoring function $s^{\theta_s}:\mathbb{R}^d\times\mathbb{R}^d\times\mathbb{R}^d\rightarrow\mathbb{R}$ (parameterized by $\theta_s$) that receives $u_i, \phi_j$ and $v_j$ as input and outputs an affinity score (scalar).
In this work, $s^{\theta_s}$ is parameterized by $\theta_s=\{W_2,W_1,W_0,r_2,r_1,r_0\}$ as follows:
\begin{equation}
    s^{\theta_s}(u_i,v_j,\phi_j)=W_2h^1_{ij}+r_2,
\end{equation}
where
\begin{align}
\label{eq:network-arch}
 h^1_{ij}=&\text{ReLU}(W_1h^0_{ij}+r_1), \nonumber \\ 
 h^0_{ij}=&[u_i,q^1_j]^T, \nonumber \\  
 q^1_j=&\text{ReLU}(W_0q^0_j+r_0), \nonumber \\
 q^0_j=&[v_j,\phi_j]^T,
\end{align}
with $W_2\in\mathbb{R}^{1\times d}, W_1,W_0\in \mathbb{R}^{d\times 2d}$, $r_1,r_0\in\mathbb{R}^d$ and $r_2\in\mathbb{R}$.
Hence, $s^{\theta_s}$ is a neural network with two ReLU activated hidden layers: The first hidden layer produces an item vector $q^1_j$ that combines the CF and CB information of the item $j$. The second hidden layer combines the CF user vector $u_i$ with $q^1_j$ to a single representation $h^1_{ij}$ that is finally transformed to a score via a linear classifier. For simplicity, we denote $s \triangleq s^{\theta_s}$.
Finally, the likelihood of a user $i$ to like (or dislike) an item $j$ is given by:
\begin{equation}
\label{eq:simple-likelihood}
    p(y_{ij}|u_i,v_j,\phi_j,\theta_s)=\sigma(y_{ij}s(u_i,v_j,\phi_j)),
\end{equation}
where $\sigma(z)\triangleq\frac{1}{1+\exp(-z)}$ is the logistic function. 

\begin{figure*}[t]
\begin{center}
\includegraphics[width=0.99\linewidth]{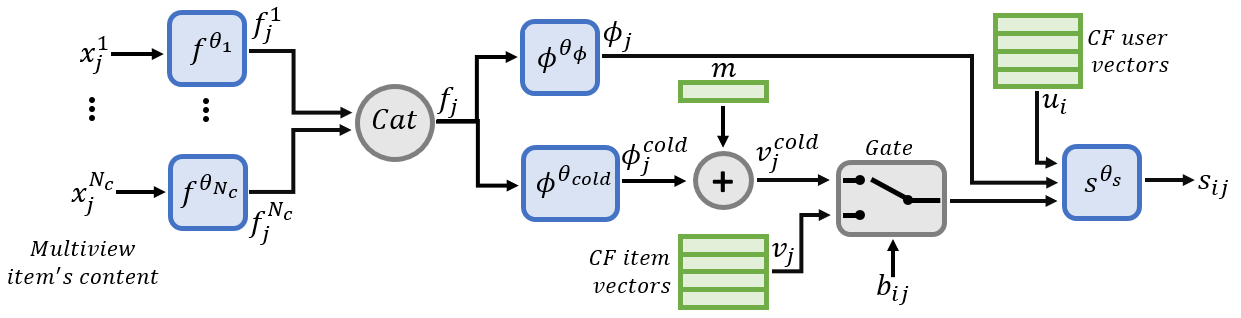}
% \vspace{-5mm}
\caption{A schematic illustration of the CWH model. $u_i$ and $v_j$ are the learned user and item CF vectors, respectively. Each content type $x_j^k$ (associated with the item $j$) is passed though a corresponding content analyzer network that encodes it as a vector $f_j^k$. The multiview content encoding $f_j$ is processed through $\phi^{\theta_{\phi}}$ and $\phi^{\theta_{cold}}$ to produce the CB representation $\phi_j$ and the CF compensation $v_j^{cold}=m+\phi_j^{cold}$, respectively. 
During training, for each training example $(i,j)$, $b_{ij}$ is repeatedly sampled from a Bernoulli distribution with a success probability $p_b(\gamma,c_j)$ that is determined by a hyperparamter (tunable knob) $\gamma \in[0,1]$ and $c_j$ - the (normalized) popularity of the item $j$ (a detailed explanation is provided in Section~\ref{subsec:p-b}).
If $b_{ij}=0$, the model behaves as a regular hybrid model passing the CF item representation $v_j$ through the stochastic gate as input to the subsequent neural scoring function $s^{\theta_s}$. Otherwise, $b_{ij}=1$ and $v_j^{cold}$ passes through the stochastic gate (instead of $v_j$), simulating a cold start scenario. In the inference phase, we compute the odds of user $i$ to like an item $j$ by setting $b_{ij}=0$ for a warm item ($j\in\mathcal{I}$) or $b_{ij}=1$ for a cold item ($j\notin\mathcal{I}$). The reader is referred to Section~\ref{sec:model} for further details. 
}
% \vspace{-4mm}
\label{fig:model_graphs}
%\vspace{-0.25cm}
\end{center}
\end{figure*}

\vspace{5mm}
\subsection{Cold-Warm Harmonization}
\label{subsec:cwh}
The CWH model is not a standard hybrid model. It includes a novel mechanism that \emph{simulates} completely cold items and forces the model to fully utilize the CB information in cases where the CF information is not available (as in completely cold items). Next, we explain the challenge of cold-warm harmonization and describe in detail our solution. 

So far, the model's objective, as defined in Equation~\ref{eq:simple-likelihood}, is inline with many previous hybrid models. However, it poses a challenge when dealing with a completely cold item $a$. In this case, the item $a$ is associated with the content representation $\phi_a$ only, while the CF representation $v_a$ is missing. This is a common scenario in real world practice, and one that we particularly want to address in scope of this work.

Since the model never actually encounters completely cold examples during training (by definition), it cannot adapt to this case. In other words, since completely cold items do not appear in the training data, the model is never actually required to use the content representation $\phi_a$ alone. Instead, the model treats the content representation $\phi_a$ as a mere `correction' over the CF representation $v_a$, and when $v_a$ is missing, the item's representation is incomplete. 
In Section~\ref{sec:results}, we investigate this case, and show that without proper treatment, the results on cold items are sub-optimal.

In order to alleviate the aforementioned problem, we propose the novel Cold-Warm Harmonization (CWH) mechanism that simulates cold-start scenarios \emph{during} the training phase. To this end, we introduce the novel CWH likelihood:
\begin{equation}
\label{eq:xxx-likelihood}
\begin{split}
    p(y_{ij}|u_i,v_j,v_j^{cold},b_{ij},\phi_j,\theta_s)=\\
            \sigma(y_{ij}s(u_i,(1-b_{ij})v_j + & b_{ij}v_j^{cold},\phi_j)).
\end{split}
\end{equation}
The CWH likelihood in Equation~\ref{eq:xxx-likelihood} introduces two new terms, $b_{ij}$ and $v_j^{cold}$ as follows: 
\begin{itemize}
    \item $b_{ij}$ - is a stochastic gate, based on an observed Bernoulli variable, that determines the likelihood state of the item as either \emph{warm} or \emph{cold}. 
    At each iteration, $b_{ij}$ is resampled from a Bernoulli distribution. 
    In the \emph{warm} state ($b_{ij}=0$), the likelihood falls back to Equation~\ref{eq:simple-likelihood}. However, in the \emph{cold} state ($b_{ij}=1$), the likelihood simulates a cold start scenario, where the term $v_j^{cold}\in\mathbb{R}^d$ replaces the missing CF representation $v_j$.
    In Section~\ref{subsec:p-b}, we elaborate on the distribution of $b_{ij}$ and propose a concrete scheme for tuning its parameter (success probability) based on the popularity distribution of warm items.

\item {\bf $v_j^{cold}$} - is the summation: $v_j^{cold}=m+\phi^{cold}_j$, where $\phi^{cold}_j\triangleq\phi^{\theta_{cold}}(f_j)$, and
    $\phi^{\theta_{cold}}:\mathbb{R}^{d_\phi}\rightarrow\mathbb{R}^{d}$ is a neural network with an identical architecture as $\phi^{\theta_{\phi}}$, but parameterized by a different set of (learned) parameters $\theta_{cold}$. $m\in\mathbb{R}^d$ is a global learned embedding vector (independent of $j$) that can be seen as a global positional bias.
\end{itemize}
The role of $b_{ij}$ is to expose the model to \emph{fake} completely cold items during training. In this case ($b_{ij}=1$), $v_j^{cold}$ is used instead of the CF representation $v_j$, ensuring the model learns a CF compensation based on the items' content. 
It is important to clarify that $\phi^{\theta_{\phi}}$ and $\phi^{\theta_{cold}}$ play different roles: $\phi^{\theta_{\phi}}$ is trained to produce $\phi_j$ - a CB vector which enhances the learned CF representation with \emph{complementary} CB information (via the concatenation $q^0_j$ in Equation~\ref{eq:network-arch}). On the other hand, $\phi^{\theta_{cold}}$ together with $m$ are trained to \emph{replace} the missing CF representation in the cold start scenario. Then, in the inference phase, when a completely cold item $a$ is introduced to the system, the role of $\phi^{cold}_a$ is to replace and compensate for the missing CF representation $v_a$ based on its content data $\mathbf{X}_a$. 

The combination of the gate $b_{ij}$, together with the network $\phi^{cold}_j$, forms a novel architecture and a key contribution of CWH that alleviates the aforementioned conflicting roles of the CB data as well as the discrepancy between training and inference in hybrid recommenders. A schematic illustration of the CWH model is depicted in Figure~\ref{fig:model_graphs}. 
\vspace{2mm}
\subsection{Popularity based Cold-Start Exposure Rate}
\label{subsec:p-b}
In many collaborative filtering datasets, items exhibit a power law distribution in which few popular items account for most of the user-items interactions in the dataset. As a consequence, the model's exposure to specific types of content is imbalanced as well. For example, consider the MovieLens dataset \cite{movielens} from Section \ref{sec:results}. 
One type of content metadata is the set of actors participating in each movie. Naturally, popular actors are mostly associated with popular movies. As a result, the model's exposure to actors is imbalanced: less popular actors are rarely introduced to the model and the model's ability to learn their CF compensation via $v^{cold}_j$ is limited. 

\begin{figure}[t]
\centering
\includegraphics[width=0.9\linewidth]{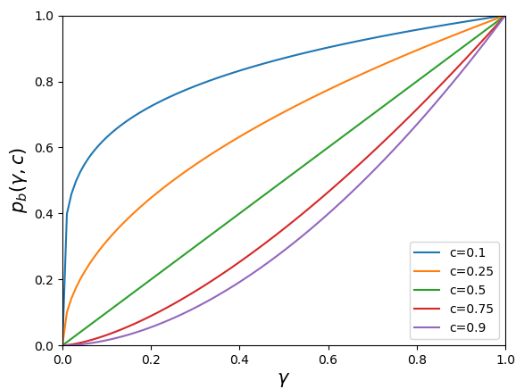}
\caption{$p_{b} (\gamma,c) = \gamma^{2c}$ for several popularity scores $c$.}
\label{fig:pb}
\end{figure}

To mitigate this problem, we propose to suppress the probability of simulated \emph{fake} cold-start scenarios for popular items, but enhance it in the case of rare items.
By taking this approach, we equipoise the model's exposure to types of content as follows: 
Denote the normalized popularity score of item $j$ by $0\leq c_j \leq 1$, where $c_{\text{max}}=1$ is associated with the most popular item, and $c_{\text{min}}=0$ is associated with the least popular item, e.g., by employing min-max normalization. 
Then, we set the parameter of the Bernoulli variable $b_{ij}$ to be $p_{b}(\gamma,c_j) = \gamma^{2c_j}$.
Figure~\ref{fig:pb} depicts $p_{b}(\gamma,c_j)$ for different popularity scores. 
We see that popular items with popularity score of $c_j>\frac{1}{2}$ produce a convex behaviour of $p_{b} (\gamma,c_j)$ with respect to the control knob $\gamma$, while rare items with popularity score $c_j<\frac{1}{2}$ produce a concave behaviour of $p_{b} (\gamma,c_j)$ with respect to $\gamma$. As a consequence, the probability to use $v_j^{cold}$ for an item $j$ with popularity score of $c_j>\frac{1}{2}$ ($c_j<\frac{1}{2}$) would be less (greater) than $\gamma$.

We believe that a careful selection of $p_{b}$ can be beneficial for the model's ability to learn CF compensation via $v^{cold}_j$. However, further investigation of the Bernoulli parameter $p_{b}$ is left outside the scope of this work and reserved for future work.

\vspace{2mm}
\subsection{Optimization and Inference}
We denote $\mathbf{B}=\{b_{ij}|(i,j)\in\mathcal{I}\times\mathcal{J}\}$ and $\bm{\Theta}=\{\mathbf{U,V},m,\theta_{CB},\theta_{\phi},\theta_{cold},\theta_s\}$. Then, by assuming normal priors over the unobserved model variables, we can write the negative log joint distribution as follows:

\begingroup\makeatletter\def\f@size{10}\check@mathfonts

\begin{equation}
\begin{split}
    \mathcal{L}&=-\log p(\mathbf{Y},\bm{\Theta}|\mathbf{B,X}) = 
    -\log\left[ p(\mathbf{Y}|\bm{\Theta},\mathbf{B,X})p(\bm{\Theta})\right] \\&= 
   -\sum_{(i,j)\in \mathcal{I}\times\mathcal{J}}\log\left[\sigma(y_{ij}s(u_i,(1-b_{ij})v_j+b_{ij}v_j^{cold},\phi_j)\right]  \\&\qquad
   +\frac{\tau}{2}\left[||\theta_s||^2_2+||\theta_{cold}||^2_2+||\theta_{\phi}||^2_2+\sum_{i=1}^{N_u}||u_i||^2_2 \right.\\&\qquad\left.\quad\quad\quad+\sum_{j=1}^{N_v}||v_j||^2_2+\sum_{k=1}^{N_c}||\theta_k||^2_2\right]+\text{const},
\end{split}
\label{eq:joint}
\end{equation}
where $\tau$ is the precision hyperparamter that controls the strength of the normal prior (similar to $L_2$ regularization). 

In practice, the negative examples ($(i,j)\notin I_y$) that appear in the likelihood term in Equation~\ref{eq:joint} are sampled in a stochastic manner, according to the procedure in \cite{barkan2016item2vec}.
We propose a \emph{Maximum A-Posteriori} (MAP) estimation, which is equivalent to the minimization of $\mathcal{L}$ w.r.t. the unobserved variables, i.e., $\bm{\Theta}^*=\argmin_{\bm{\Theta}} \mathcal{L}$, where the optimization is carried out using stochastic gradient descent.
 
At inference, we compute the odds of user $i$ to like an item $j$ following Equation~\ref{eq:xxx-likelihood} by setting $y_{ij}=1$,  with $b_{ij}=0$ or $b_{ij}=1$, if $j$ is warm ($j\in\mathcal{J}$) or cold ($j\notin\mathcal{J}$), respectively. 
\vspace{3mm}

\section{Experimental Setup and Results}
\label{sec:results}
We present experimental evaluation, consisting of two major parts:
In the first part (\textbf{P1}), we evaluate the capabilities of CWH in handling completely cold-start scenarios (new items that were just introduced to the system and hence don't have any usage data). We consider a case in which several completely cold items are integrated into an existing warm catalog of items. We focus on the inherent trade-off that arises when integrating warm and cold items together and demonstrate the ability of CWH to gently balance between the two objectives:  preserving the performance on the warm catalog and promoting the items from the new (completely cold) catalog. 

The second part of the evaluation (\textbf{P2}) demonstrates the performance of CWH as a hybrid recommender system utilizing both usage data (implicit ratings) as well as the content based data that exist for the items.

\vspace{2mm}
\subsection{Experimental setup}
\label{subsec:exp-setup}
Our evaluations are based on a classical user-item prediction task, i.e. the ability to recommend the correct item to the right user. Our datasets consist of users and their lists of items (items purchased or consumed by each user). In what follows, we describe the datasets that are used in this research. 

\vspace{2mm}
\subsubsection{Datasets}
\hfill\\
Three datasets from different domains are considered: 
\begin{itemize}
    \item \textbf{\emph{Movies}}: This dataset is based on the public MovieLens dataset \cite{movielens}. It consists of $22M$ ratings for $N_v=34K$ movies by $N_u=247K$ users. Each user-movie interaction is rated by a 5-star scale, and we considered 3.5 stars and above as a positive signal.
    Importantly, we enriched the movies with content metadata we collected from IMDB\footnote{\url{www.imdb.com}}. The metadata for each movie consists of textual, categorical and numerical data fields as follows: the plot, the list of actors, the director, a list of genres, languages, the year of release and additional descriptive tags. We also extracted the poster image for each movie as a visual content data.

    \item
    \textbf{\emph{CiteULike}}: This dataset is based on \cite{citeulike-ref} - a public collection of users' reading lists crawled from the CiteULike website taken from \cite{wang2015collaborative}. It consists of $205K$ user-item interactions of $N_v=17K$ articles that were read by $N_u=5.5K$ users (we used the CiteULike-a dense fold as in \cite{wang2015collaborative}).
    The metadata for each article consists of the article's title, the abstract, and a list of descriptive tags.

    \item 
    \textbf{\emph{Apps}}: This is a propriety dataset that was collected from the Microsoft Windows Store. The dataset consists of $5M$ anonymous user sessions, where each session consists of a list of applications (apps) that were co-clicked during the same browsing session. It contains $20M$ user-item  interactions and $N_v=33K$ unique apps. The apps metadata consists of the app's title, its description, the release date, a list of descriptive tags, and the app's icon image (as visual content).
\end{itemize}

\vspace{3mm}
\subsubsection{Train / Test Partitioning} \hfill\\
\label{subsubsec:partitioning}
We now turn to describe the train / test split employed on each of the above datasets. 
We considered users with at least $8$ items. For each user, we randomly drew two items to form the \emph{test} set and another two items to form the \emph{validation} set. 
Then, we choose 20\% of the items and removed all their interactions from the training set in order to simulate cold items. Half of these items (10\% of the 20\%), were used in the validation set and the second half were used for the test set. Cold items that were selected for the validation set, were removed from the test set and vice versa.

The \emph{warm} and \emph{cold} items in the test set were used for the first part (\textbf{P1}) of the evaluations, while in the second part (\textbf{P2}) of the evaluation we used the \emph{warm} items without the (completely) \emph{cold} items that do not appear in the training set. Hence, the second part (\textbf{P2}) of the evaluation is comparable to standard evaluation of recommender system. 
The validation set was used to tune the model's hyperparameters. 
We run each experiment 10 times with different realizations of the train / validation / test partition and report the mean results.

Special care is required in the selection process of the simulated cold items. 
In a recent review and evaluation of modern recommendation algorithms, Dacrema \etal showed that the popularity distribution of test items has a significant impact on a model's performance evaluation (see Section 3.6 in~\cite{are_we_progress}). This finding stems from the fact that the items' content distribution \emph{cannot} be considered statistically independent of their popularity. 
In other words, popular items exhibit different content distribution than the content distribution of rare items.
Hence, without proper selection of cold items, the evaluations would be inaccurate, inconsistent and irreproducible in real-world, especially in datasets with a small number of items or a high popularity skew~\cite{are_we_progress}. 

When new items are introduced to the system, these items are cold by definition. However, we should not assume that these items will remain unpopular in the long run. If we concentrate our evaluations on the unpopular items (similar to many previous works), our results will not reflect the actual business scenario at hand: introducing new items that may, with time, become popular. 
In other words, a real-world recommender needs to handle different cold (new) items, some of which are expected to become popular in the future while others will remain unpopular also in the long run. 
Therefore, the popularity of items in the test set should follow that of regular items in the training set. This is in contrast to other works that did not take this consideration into account~\cite{are_we_progress}. 

In order to guarantee similar popularity distributions between the train, test and validation sets, we devised the following procedure: (1) We sort the items according to their popularity. Then, (2) we select each tenth item for the test set and its successive item for the validation set. The rest of the items consist the train set (the ratios can be adjusted as needed). Different folds are obtained by considering different offsets.
The resulting train, test and validation sets consist of cold items that exhibit the same popularity distribution as new items in a real-world scenario.

\vspace{3mm}
\subsubsection{Evaluation Measures}
\hfill\\
Using the test sets described above, the following measures are reported:

\begin{itemize}
    \item \textbf{\emph{Mean Hit Rate at K (HR@K)}}: For a user-item pair $(i,j)$, Hit Rate at K outputs $1$ if the target item $j$ is ranked among the top $K$ recommendations w.r.t. to the user $i$, otherwise $0$. We report the average HR@K across all test pairs. 
    \item \textbf{\emph{Mean Reciprocal Rank at K (MRR@K)}}: This measure computes the reciprocal rank of the target item $j$ among the top $K$ recommendations w.r.t. to the user $i$. If the rank of the item $j$ is not the top $K$ recommendations, the result is $0$. Finally, MRR@K is obtained by averaging the reciprocal rank over the entire test set of user-item pairs.
\end{itemize}

We refer the reader to \cite{metrics-ref} for a broader discussion and the motivation behind these evaluation measures.

\begin{figure*}[t]
\centering
\begin{tabular}{c}
    \includegraphics[scale=0.4]{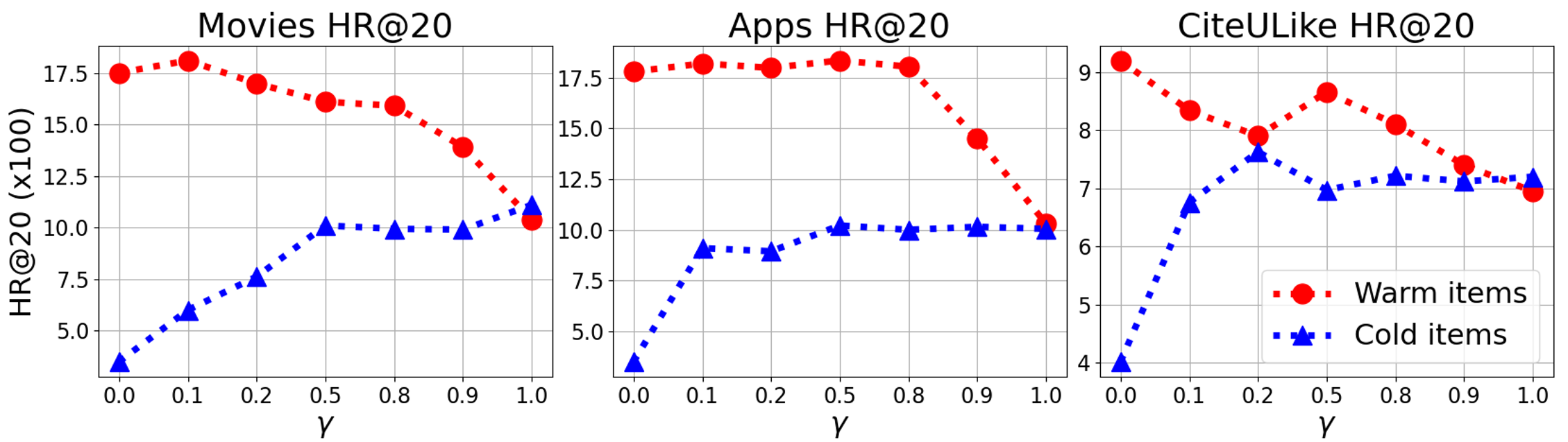}
\end{tabular}
% \vspace{-2mm}
\caption{HR@20 (x100) for different exposure levels ($\gamma$) of $v_j^{cold}$ for the Movies (left), Apps (center) and  CiteULike (right) datasets.}
\label{fig:apps-hr}%\vspace{-3mm}
\end{figure*}

\vspace{3mm}
\subsubsection{Baselines and Hyperparameters Configuration}
\hfill\\
All hyperparameters were set based on the separate validation set (described in the beginning of Section \ref{subsec:exp-setup}). All the models were trained with an early stopping procedure, based on the monitored measures (utilizing the same validation set). We evaluated the following models:
\begin{itemize}
    \item \textbf{\emph{CWH (this paper)}}: The Cold-Warm Harmonization model from Section~\ref{sec:model}.
We used three types of content analyzers, each analyzes a different type of content data:
\begin{itemize}
\item For categorical data fields and tags we used a linear neural network $f^{\theta_{tags}}$ that encodes the category / tag into a 100 dimensional vector. 
\item For textual description, we used the pre-trained BERT-Base model from~\cite{barkan2019scalable}, denoted by $f^{\theta_{text}}$. Specifically, for each item, we considered the first 512 words in its textual description. Then, we computed an average across the hidden tokens in the last hidden layer to produce one 768 dimensional vector. We then passed this vector to an additional linear projection layer which yields a final 100 dimensional description representation vector.
\item For visual content, we used the backbone of a pre-trained ResNet18 network \cite{he2016deep}, denoted by $f^{\theta_{image}}$. We applied another linear layer on top of the output of the ResNet18 to produce a final 100 dimensional image representation vector. 
\end{itemize}

Finally, we concatenated the outputs from $f^{\theta_{tags}}, f^{\theta_{text}}, f^{\theta_{image}}$ to form a 300 dimensional vector as the input to the multi-view content analyzer function $\phi^{\theta_{\phi}}$ and to the CF compensation function $\phi^{\theta_{cold}}$. The output dimensions of $\phi^{\theta_{\phi}}$ and $\phi^{\theta_{cold}}$ were set $d=100$, which matches the CF dimension as well.

CWH was optimized by using a default Adam~\cite{kingma2014adam} optimizer with a batch size of 32. The regularization hyperparameter was set to $\tau=1e-5$ (based on the validation set). Convergence was observed after 50-60 epochs, depending on the dataset.

\item \textbf{\emph{CF-only}}: This model is an ablated version of CWH that uses its CF component only. The architecture of the CF component is based on the Neural Collaborative Filtering (NCF) model from \cite{ncf}. This baseline is a pure CF model, hence unable to utilize content and cannot support cold item recommendations. Therefore, results for CF-only are reported in the second part (\textbf{P2}) of the evaluation only.

\item
\textbf{\emph{CB-only}}: This model is an ablated version of CWH that does not learn the CF item representations. The CF items' representations: $\mathbf{V}, v_j^{cold}$ and $m$ are effectively set to zero. It uses the same multiview content embedding function $\phi^{\theta_{\phi}}$ and the same objective function as CWH.

\item \textbf{\emph{CB2CF}}: This is the model from \cite{cb2cf}. CB2CF employs a deep regression model that learns a mapping from items CB representations to their corresponding CF representations. This mapping is later used in order to estimate the CF representations for cold items based on their content.

\item \textbf{\emph{ItemKNN-CFCBF}}: This is a simple hybrid baseline from \cite{are_we_progress} %Sec.~\ref{related-Work}) 
that was recently shown to outperform multiple state-of-the-art hybrid models (e.g., CVAE \cite{cvae}). 
The original version of ItemKNN-CFCBF from \cite{are_we_progress} does not support completely cold items since it assumes the existence of a CF representation.
Hence, in order to extend the applicability of ItemKNN-CFCBF to cold items, we replace set their CF representations by a weighted average of the CF representations of their nearest warm items, where the affinity between the cold item to the warm items (and hence the weighting) is determined by the cosine similarity between the CB representations.
\end{itemize}

For fairness, all the evaluated models (except for CF-only) utilized the same multiview CB data as well as the same pre-trained models as backbones (BERT, ResNet).

\begin{table*}[t!]
\centering
\caption{\label{tab:results} MRR@20 (x100) for the warm, cold and unified items sets. CWH results reported for the optimal gamma ($\gamma^*$) values of 0.6, 0.2, and 0.6 for the Apps, CiteULike, and Movies datasets, respectively. ItemKNN stands for ItemKNN-CFCBF}
% \vspace{-2mm}
\normalsize

\begin{tabular}{|c|c||c|c|c|c|}
\hline \textbf{Dataset} & \textbf{Test Set} & \textbf{CB-only} & \textbf{CB2CF} & \textbf{ItemKNN} & \textbf{CWH} \\ \hline\hline

\multirow{3}{*}{\rotatebox[origin=c]{0}{\textbf{Apps}}} 
&\textbf{Warm} & 2.697 & 2.667 & 1.906 & \textbf{3.371} \\ \cline{2-6}
&\textbf{Cold} & 2.941 & 1.421 & 1.281 & \textbf{3.291} \\ \cline{2-6}
&\textbf{Unified} & 2.988 & 2.044 & 1.594 &\textbf{3.308}\\ \cline{2-6}
\hline \hline

\multirow{3}{*}{\rotatebox[origin=c]{0}{\textbf{CiteULike}}}  
&\textbf{Warm} & 1.417 & 0.928 & 0.817 & \textbf{1.780} \\ \cline{2-6}
&\textbf{Cold} & 0.834 & 1.101 & 0.894 & \textbf{3.049} \\ \cline{2-6}
&\textbf{Unified} & 1.125 & 1.015 & 0.855 & \textbf{2.414} \\ \cline{2-6}
\hline \hline

\multirow{3}{*}{\rotatebox[origin=c]{0}{\textbf{Movies}}}
&\textbf{Warm} & 1.306 & 2.639 & 2.686 & \textbf{2.962} \\ \cline{2-6}
&\textbf{Cold} & 1.835 & 1.148 & 1.470 & \textbf{2.418} \\ \cline{2-6}
&\textbf{Unified} & 1.581 & 1.863 & 2.078 & \textbf{2.691} \\ \cline{2-6}
\hline

\end{tabular}
\vspace{5mm}
\end{table*}

\begin{figure*}[t]
    %\vspace{-0.25cm}
    \centering
    % \begin{center}
    \begin{tabular}{c}
        {\includegraphics[scale = 0.42]{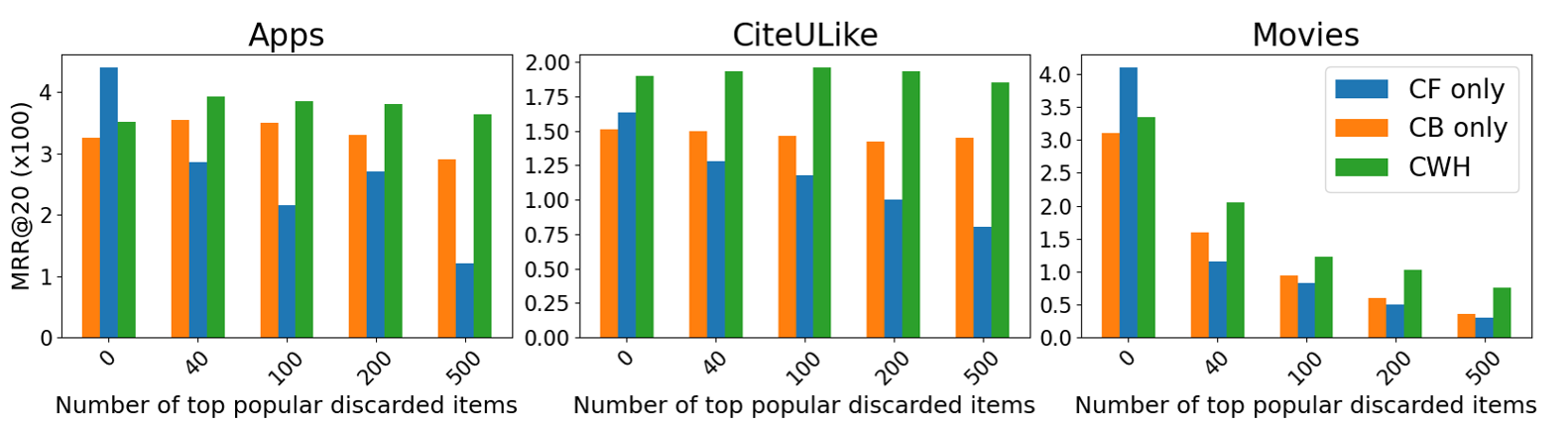}}
    \end{tabular}
    % \vspace{-3mm}
    % \end{center}
    \caption{
    MRR@20 (x100) results for different combinations of dataset and methods, across different popularity regimes. See Section \ref{subsec:res_longtail} for details.
    }\vspace{5mm}
    \label{fig:pop_graphs_movies}
\end{figure*}

\vspace{1mm}
\subsection{Warm vs. Cold Balancing (P1)}
\label{subsec:res_tradeoff}
We begin by evaluating the ability of CWH to conduct an adaptable integration of multiple cold items into an existing model (of warm items) and produce recommendation lists that include both cold and warm items. This ability is determined by the $\gamma$ hyperparameter that serves as a knob to control the exposure rate of the  $v_j^{cold}$ token at the training phase and has a significant effect on the results of CWH (note that we set $c_j=0.5$ for all $j\in\mathcal{J}$).

Therefore, when $\gamma=0$, the model is never exposed to cold items and acts as an ordinary hybrid model. Similar to previous hybrid models, in this case, CWH is focused on warm item recommendations. While it manages to produce cold item representations based on its content analyzers, these representations are sub-optimal. 
On the other hand, when $\gamma=1$, the model is exposed to cold items only. As a consequence, it cannot learn CF item representations and collapses to a pure CB model. In this case, CWH is agnostic to the fact that an item is warm or cold since it considers its content only.

As we show next, an insufficient exposure of the model to cold items hinders the model's ability to generalize for such items. In contrast, over exposure to cold items hinders the model's performance for the warm items. 
This trade-off is clearly noticeable in Fig.~\ref{fig:apps-hr}, where we present the HR@20 of warm and cold items for different exposure levels ($0\leq \gamma \leq 1$). The results in Fig.~\ref{fig:apps-hr} align with the aforementioned theory. Specifically, one can notice that when $\gamma=1$, the warm and cold lines coincide (up to an empirical variance). These results indicate that CWH provides the ability to balance between warm and cold items according to varying business needs.

Next, we turn to compare CWH against the other baselines. Unlike CWH, these baselines do not feature a control-knob to smoothly balance between warm and cold items. For each method we report the evaluation measurements for warm and cold items separately. In addition, we consider the case of a unified test set consisting of $90\%$ warm items and $10\%$ cold items. This, we believe, corresponds to a realistic scenario in which the warm catalog is extended using a new cold catalog which is $\frac{1}{9}$ times its size. 

The results for Movies, CiteULike, and Apps are presented in Table~\ref{tab:results}. For CWH, we report the results for $\gamma^{*}$, i.e., the $\gamma$ value that maximizes the $MRR@20$ on the unified catalog on each of the validation sets. We see that CWH outperforms the baselines across all datasets and test sets (warm, cold, and unified).

\vspace{1mm}
\subsection{Warm Items Analysis (P2)} \label{subsec:res_longtail}
In this section, we focus on evaluating CWH as a hybrid recommender on warm items only i.e., items that were present in the training set. 
Some of the warm items are highly popular, while others are extremely rare with an insufficient number of observations a.k.a ``long tail'' items. As a consequence, CF models suffer from performance degradation for items within the ``long tail'' \cite{park2008long,vbn,barkan-etal-2020-bayesian,barkan2017bayesian}.
A hybrid model is expected to compensate for the lack of data in rare items by utilizing the available content data. 

In order to evaluate this capability, we perform a careful analysis according to the amount of available data (popularity).
We denote the test set of user-item pairs as $T= \{(i,j)\}$ and denote the set of the $r$ most popular items as $A_r$. 
Then, we define $T_r = \{(i,j): j \notin A_r \}$, \ie  $T_r$ as a subset of $T$ in which excludes the top $r$ most popular items. We compute the MRR@20 for each $T_r$, where $r \in \{ 0, 40, 100, 200, 500\}$.

The results for the Movies, CiteULike, and Apps datasets are depicted in Fig.~\ref{fig:pop_graphs_movies}. For the sake of clarity, we omit ItemKNN-CFCBF and CB2CF as they perform the worst. Yet, we do include CB-Only for ablation purposes and the CF-only baseline for examining the behaviour of a CF system across the different popularity regimes. This analysis reveals different ``layers'' that are usually concealed due to the predominance of popular items. 
Figure~\ref{fig:pop_graphs_movies} shows that CWH outperforms the other baselines in the majority of popularity values. 
Hence, we conclude that CWH attains excellent results as a regular hybrid recommender (in addition to its ability to integrate completely cold items to an existing catalog).
\vspace{2mm}

\section{Conclusion}
\label{sec:conclusions}
We discuss a common scenario in which a recommender system is expected to perform accurately both on warm items as well as on completely cold items.
To this end, we present the CWH model - a deep hybrid model that employs a novel stochastic gating mechanism specifically designed to mitigate the aforementioned difficulty. Through a tunable knob, CWH allows to find the desired balance according to business needs. 
The effectiveness of CWH is demonstrated on three datasets, where it is shown to outperform other alternatives.
\vspace{4mm}

\bibliography{references}
\bibliographystyle{IEEEtran}

\end{document}